# Diverging evolution of light pollution indicators: can the Globe at Night and VIIRS-DNB measurements be reconciled?

Salvador Bará[1,*], and José J. Castro-Torres[2]

[1] *Independent scholar. Former profesor titular (retired) at Universidade de Santiago de Compostela (USC), Santiago de Compostela, 15782 Galicia (Spain, European Union)*

[2] *Laboratory of Vision Sciences and Applications, Department of Optics, University of Granada, Granada, 18071 (Spain, European Union)*

* Corresponding author. e-mail: salva.bara@usc.gal

**Abstract:** The radiance of nighttime artificial lights measured by the VIIRS-DNB instrument on board the satellite Suomi-NPP increases at an average rate ~2.2 %/yr worldwide, whereas the artificial radiance of the night sky deduced from the Globe at Night (GAN) unaided-eye observations of the number of visible stars is reported to increase at an average rate ~9.6 %/yr. The difference between these two estimates is remarkable. This raises the question of whether the diverging temporal evolution of these indicators could be due to changes in the spectral composition of outdoor artificial light, consequence of the current process of replacement of lighting technologies. This paper presents a model for evaluating the temporal rate of change of different light pollution indicators and applies it to the VIIRS-DNB vs GAN issue, based on available data. The results show that the reported difference could be explained by spectral changes alone, if the visual GAN observations are made with scotopic or mesopic adaptation at definite times under some particular transition conditions. In case of photopic adapted observers, however, reconciling these two measurement sets requires the existence of GAN-specific light sources that affect the Globe at Night observations but do not show up in the VIIRS-DNB data. The lumen emissions of these GAN-specific sources for photopic observers should increase at a rate larger than 9%/yr worldwide.

## 1. Introduction

How fast are artificial light emissions growing worldwide? This is one of the open questions in light pollution research, prompted by the rising awareness of the unwanted side-effects of artificial light at night. One approach to address this issue is to analyze the temporal evolution of light pollution indicators. Light pollution indicators are observable quantities derived from the spectral radiance of the anthropogenic light [1-4]. They can be linear or non-linear, depending on their definition in terms of the radiance at the observer location [5,6]. They can refer to the light received directly from the light sources, to the light scattered by small-scale inhomogeneities of the material medium through which light propagates (e.g. the atmosphere or the water column), or to a combination of both.

Light pollution indicators are evaluated in different spectral bands, depending on the effects under study and the available measurement devices. Common detection bands include, among others, the human photopic and scotopic spectral luminous efficiencies [7,8], the five human photoreceptor alfa-opic curves [9-11], various sets of classical astronomical filters [12] including the RGB system [13-15], satellite [16,17] and ground-based [18,19] radiometer passbands, and a wide range of animal photoreceptor sensitivities and action spectra [20].

Light pollution indicators may provide valuable insights about the evolution of the global amount of artificial light emissions. The rationale is that the radiance measured at the observer location depends linearly on the weighted emissions of the light sources, mediated by the state of the atmosphere [2,5,21,22]. It has long been known, however, that the time evolution of different indicators, even if correlated, may differ in magnitude and sign. The same change in the number and types of artificial light sources and/or in their geographical distribution around the observer may result in increasing values for some indicators and decreasing values for others. A situation of this kind typically arises when there is a shift in the spectral composition of the light and the indicators are evaluated in different spectral sensitivity bands [23,24]. Different rates of progression of the indicator values can also be experienced, even in the same band, if the angular distribution of the emitted radiance does change in time (e.g. by changes in the shielding or the aiming of the lights). While the evolution of each individual indicator provides information about the changes in the specific effect it measures (which may well be different from one indicator to another), their divergent collective behavior can be a problem when it comes to using them as a proxy to assess other quantities of interest as e.g. the evolution of the global amount of artificial light emissions.

An interesting case of diverging temporal evolution has been recently reported [25]. According to these results, the artificial night sky brightness derived from the Globe at Night unaided-eye star-counting citizen science campaigns increased at an average rate $9.6\pm0.4$ %/yr during the period 2011-2022, substantially faster than the estimated ~2.2 %/yr average rate of increase of the overall artificial light emissions derived from the VIIRS-DNB band satellite radiance measurements in the period 2012-2016 [26].

Can these results be reconciled? If so, what can they teach us about the processes at play? Recall that, everything else being constant (average state of the atmosphere, spatial distribution, and spectral and angular emission patterns of the ensemble of light sources), one would expect that the rate of change of the artificial night sky brightness should be equal to the rate of change of the total (distance-weighted) light emissions in the territory surrounding the observer, irrespective from the spectral sensitivity bands in which each indicator is measured. Different rates of change could arise, however, if either the average state of the atmosphere, the characteristics of the sources (spatial distribution and spectral and angular emission patterns), or a combination of these factors, would actually change in time.

This paper describes a model for analyzing these situations, derived from first principles of light pollution propagation. This approach is applied to interpret the seemingly divergent results from Globe at Night observations [25] and VIIRS-DNB satellite nighttime measurements [26]. The goal is to assess whether their different evolution rates can be explained by the change of spectrum of artificial light associated with the current transition from low CCT (Correlated Color Temperature) high-pressure sodium lamps to high CCT LED (Light Emitting Diode) sources, or other additional factors have to be taken into account in order to fully explain them.

## 2. Time evolution of light pollution indicators

The two light pollution indicators analyzed in this paper are (i) the artificial radiance of the night sky, $L_{\text{GAN}}(t)$, estimated from multiple unaided-eye observations of the number of visible stars on selected regions of the sky, made by Globe at Night citizen scientists in a wide set of locations worldwide, and (ii) the top-of-atmosphere radiance of nighttime artificial lights, $L_{\text{DNB}}(t)$, measured by the VIIRS radiometer onboard the Suomi-NPP satellite in its day-night band (DNB). The basic equations for their time evolution are described in this section.

From a formal standpoint both indicators are angular field-of-view averages of spectral radiances integrated within their respective sensitivity bands. The angular averaging for $L_{\text{GAN}}(t)$ is made over the solid angle span of the selected sky region (typically a well-known constellation or asterism), being the spectral integration carried out within the human photopic, scotopic, or mesopic bands, depending on the state of adaptation of the observer. The corresponding luminance of the sky, in cd·m$^{-2}$, can be obtained by multiplying the radiance $L_{\text{GAN}}(t)$ in W·m$^{-2}$·sr$^{-1}$ by the luminous efficacy of radiation, e.g. $K_{\text{r}} = 683$ lm·W$^{-1}$ for photopic observers. The angular averaging for $L_{\text{DNB}}(t)$, in turn, is carried out within each pixel (15×15 arcsecond-square) of the satellite image, being the spectral integration made in the DNB sensitivity band. For the sake of definiteness, the following equations will generically refer to the indicators as 'radiances', although all expressions in this section could equally be applied to any other linear indicator (e.g. horizontal irradiances or illuminances) by substituting the appropriate quantities.

A step-by-step derivation of the time evolution equations is developed here. The indicators will be denoted as $L_{\beta}(t)$, $\beta \in \{\text{GAN}, \text{DNB}\}$. The value of each indicator at each moment of time results from the contributions $L_{\beta}^{s}(t)$ of several sets of light sources, labeled by the superindex $s=1,...,S$.

$$L_{\beta}(t) = \sum_{s=1}^{S} L_{\beta}^{s}(t) \qquad (1)$$

By "sets of light sources" we refer here to the different categories in which the whole ensemble of light sources can be conceptually divided by practical reasons. Examples of different sets are streetlights, domestic lights, ornamental lights, LED billboards... Any particular set of sources may be composed by a mix of lamps of different technologies and with different spectra, for example HPS (High-Pressure Sodium) and LED, combined in variable proportions which may change in time. Additionally, the contribution $L_\beta^s(t)$ of each set of light sources to the value of the indicator $L_\beta(t)$ can take place through different physical processes (e.g. scattering in the atmosphere, intraocular scattering, direct propagation from ground to a satellite, and others). Denoting by $L_\beta^{sp}(t)$ the contribution of the $s$-th set of light sources to the indicator β through each process $p=1,...,P$, we can write:

$$L_\beta^s(t) = \sum_{p=1}^{P} L_\beta^{sp}(t) \qquad (2)$$

Now let us recall that any $L_\beta^{sp}(t)$ is linear on the radiant emissions of the light sources. In particular, it can be calculated as a weighted integral over wavelengths of $E_\beta^s(\lambda, t)$, the spectral radiant flux emitted per unit area averaged over the relevant region of the territory whose emissions affect the indicator value. A formal derivation of this relationship may be found in Ref. [5]. Hence we can write:

$$L_\beta^{sp}(t) = \int F_\beta^{sp}(\lambda, t) \, E_\beta^s(\lambda, t) \, d\lambda \qquad (3)$$

where $L_\beta^{sp}(t)$ has dimensions W·m$^{-2}$·sr$^{-1}$, $t$ is the time, and $E_\beta^s(\lambda, t)$ is expressed in W·m$^{-2}$·nm$^{-1}$. $E_\beta^s(\lambda, t)$ refers to the flux emitted by all sources of the s-th set. In present-day installations it is mostly sent towards the surrounding surfaces (pavements, façades, grasslands,...) and, in a smaller proportion, directly towards the upper hemisphere.

The function $F_\beta^{sp}(\lambda, t)$ in Eq. (3) has dimensions sr$^{-1}$. It accounts for the light propagation through the $p$-th process, since the light leaves the sources until it is detected. It can be formally defined as the spectral radiance at the observer location, weighted by the spectral sensitivity of the β-band, produced in the process $p$ by a unit amount of spectral radiant flux per unit area $E_\beta^s(\lambda, t)$ emitted by the $s$-th set of light sources.

$F_\beta^{sp}(\lambda, t)$ depends on the particular emission-propagation-detection configuration. It generally contains the result of many intermediate operations, often including path integrals and spatial and angular integrations over the sources' spatial distribution and angular emission patterns, respectively. It takes into account the wavelength-dependent optical properties of the surfaces with which the light interacts (bidirectional reflectance distribution functions) and the state of the atmosphere through which it propagates, which jointly determine the absorption and scattering of light from the source to the detector. The spectral sensitivity of the detector is also included in its definition. Its specific values have to be calculated for each indicator and will be discussed in section 3. The explicit form of $F_\beta^{sp}(\lambda, t)$, however, is not required for the present section.

To obtain the time evolution equations, let us assume that the $s$-th set of light sources is composed of lamps with $k = 1, ..., K$ different technologies (HPS, LEDs of various CCT, etc), whose nominal spectral radiant fluxes, normalized to one lm, are $\widehat{\Phi}_k(\lambda)$. The spectral flux emitted per unit area of the territory, $E_\beta^s(\lambda, t)$, can then be expressed as:

$$E_\beta^s(\lambda, t) = \sum_{k=1}^{K} N_{\beta k}^s(t)\, \widehat{\Phi}_k(\lambda) = N_\beta^s(t) \sum_{k=1}^{K} \gamma_{\beta k}^s(t)\, \widehat{\Phi}_k(\lambda) \qquad (4)$$

where $N_{\beta k}^s(t)$ is the number of lumen emitted per unit area by the $k$-th type of lamps of the $s$-th set, $N_\beta^s(t) = \sum_{k=1}^{K} N_{\beta k}^s(t)$ is the total lumen emitted per unit area by the $s$-th set, and $\gamma_{\beta k}^s(t) = N_{\beta k}^s(t)/N_\beta^s(t)$ is the fractional contribution of the $k$-th type of lamps to the total lumen emissions of the $s$-th set, being $\sum_{k=1}^{K} \gamma_{\beta k}^s(t) = 1$. The dimensions of $\widehat{\Phi}_k(\lambda)$ are W·m⁻²·nm⁻¹/(lm·m⁻²)= W·nm⁻¹·lm⁻¹.

Combining the above equations, we have

$$L_\beta(t) = \sum_{s=1}^{S} N_\beta^s(t) \sum_{k=1}^{K} \gamma_{\beta k}^s(t)\, \mathcal{H}_{\beta k}^s(t) \qquad (5)$$

where $\mathcal{H}_{\beta k}^s(t) \equiv \sum_{p=1}^{P} \mathcal{H}_{\beta k}^{sp}(t)$ is a sum over processes, being each term a short-hand notation for the integral

$$\mathcal{H}_{\beta k}^{sp}(t) \equiv \int F_\beta^{sp}(\lambda, t)\, \widehat{\Phi}_k(\lambda)\, d\lambda \qquad (6)$$

The $\mathcal{H}_{\beta k}^{sp}(t)$ terms depend on the spectra of the sources, the physics of light propagation, and the detection band, but not on the amount of lumen emitted per unit area, $N_\beta^s(t)$, nor on their allocation among several types of lighting technologies, $\gamma_{\beta k}^s(t)$. The dimensions of $\mathcal{H}_{\beta k}^{sp}(t)$ are, from Eq. (5), W·m⁻²·sr⁻¹/(lm·m⁻²) = W·sr⁻¹·lm⁻¹. $\mathcal{H}_{\beta k}^{sp}(t)$ can be understood as the radiance detected in the β-band per lm·m⁻² emitted by the $k$-th type of lamps of the $s$-th set of sources and propagated through the $p$-th process.

Denoting by a dot above any variable its derivative with respect to time, $\dot{x} \equiv dx/dt$, the time evolution equation of the radiance $L_\beta(t)$ in Eq. (5) is :

$$\dot{L}_\beta(t) = \sum_{s=1}^{S}\sum_{k=1}^{K} \left[ \dot{N}_\beta^s(t)\, \gamma_{\beta k}^s(t)\, \mathcal{H}_{\beta k}^s(t) + N_\beta^s(t)\, \dot{\gamma}_{\beta k}^s(t)\, \mathcal{H}_{\beta k}^s(t) + N_\beta^s(t)\, \gamma_{\beta k}^s(t)\, \dot{\mathcal{H}}_{\beta k}^s(t) \right] \qquad (7)$$

where the first term accounts for the changes in the total amount of emitted lumen per unit area via $\dot{N}_\beta^s(t)$, the second describes the effects due to the changes of the spectra of the sources via $\dot{\gamma}_{\beta k}^s(t)$, and the third one accounts for the effects of the changes in the remaining parameters of the emission-propagation-detection processes, via $\dot{\mathcal{H}}_{\beta k}^s(t)$.

The relative rate of change, in units yr⁻¹ or equivalently in %/yr, is:

$$\frac{\dot{L}_\beta(t)}{L_\beta(t)} = \frac{\sum_{s=1}^{S}\sum_{k=1}^{K} \dot{N}_\beta^s(t)\, \gamma_{\beta k}^s(t)\, \mathcal{H}_{\beta k}^s(t)}{\sum_{s=1}^{S}\sum_{k=1}^{K} N_\beta^s(t)\, \gamma_{\beta k}^s(t)\, \mathcal{H}_{\beta k}^s(t)} + \frac{\sum_{s=1}^{S}\sum_{k=1}^{K} N_\beta^s(t)\, \dot{\gamma}_{\beta k}^s(t)\, \mathcal{H}_{\beta k}^s(t)}{\sum_{s=1}^{S}\sum_{k=1}^{K} N_\beta^s(t)\, \gamma_{\beta k}^s(t)\, \mathcal{H}_{\beta k}^s(t)}$$
$$+ \frac{\sum_{s=1}^{S}\sum_{k=1}^{K} N_\beta^s(t)\, \gamma_{\beta k}^s(t)\, \dot{\mathcal{H}}_{\beta k}^s(t)}{\sum_{s=1}^{S}\sum_{k=1}^{K} N_\beta^s(t)\, \gamma_{\beta k}^s(t)\, \mathcal{H}_{\beta k}^s(t)} \quad (8)$$

where the terms in the right-hand side of Eq. (8) are the relative rates of change of the indicator due to changes in the total lumen emissions, in their allocation to different source spectra, and in the remaining parameters of the emission-propagation-detection processes, respectively.

## 3. Expected evolution of VIIRS-DNB and visual GAN radiances

### *3.1 Basic approach*

In this section we apply the formulation developed in Section 2 to analyze the divergence between the measured VIIRS-DNB and GAN rates, $\dot{L}_{\mathrm{DNB}}(t)/L_{\mathrm{DNB}}(t)$ and $\dot{L}_{\mathrm{GAN}}(t)/L_{\mathrm{GAN}}(t)$. This is a complex issue because a sizeable part of the required quantitative information is not yet available. It is possible, however, to get some insights based on known facts. These insights are expected to provide useful suggestions for further research.

Our approach is to assess the (null) hypothesis that the VIIRS-DNB and GAN measured radiances respond to the emissions of the same set of light sources. If this hypothesis is proven false, then there should be at least some set of specific set of light sources affecting one of the indicators and not the other. The basic, simplifying assumptions made for this calculation are listed here:

(i) A single set of light sources ($S$=1) for both indicators $\beta = \{\mathrm{DNB}, \mathrm{GAN}\}$

(ii) Two kinds of lamp technologies ($K$=2) in that set, namely high-pressure sodium (HPS) and light-emitting diodes (LED) of correlated color temperatures 4000 K and 3000 K (which will be studied separately), with one-lumen normalized spectra $\widehat{\Phi}_{\mathrm{HPS}}(\lambda)$ and $\widehat{\Phi}_{\mathrm{LED}}(\lambda)$, respectively.

(iii) A transition process in which an initial population of HPS lamps is progressively replaced by LED, with fractional lumen contributions $\gamma_{\mathrm{HPS}}(t)$ and $\gamma_{\mathrm{LED}}(t)$, where $\gamma_{\mathrm{LED}}(t) = 1 - \gamma_{\mathrm{HPS}}(t) \equiv \gamma(t)$, and $\gamma_{\mathrm{HPS}}(t) = 1 - \gamma(t)$.

(iv) $\mathcal{H}_{\beta k}$ factors, denoted as $\mathcal{H}_{\beta,\mathrm{HPS}}$ and $\mathcal{H}_{\beta,\mathrm{LED}}$, independent of time (the effects of the time dependence of these factors will be discussed in Section 4).

(v) A single process for DNB radiance propagation (reflection of lamps' light on pavements and other surfaces and propagation to the top of the atmosphere) and two processes for GAN (skyglow generated by the atmospheric scattering of light, and equivalent glare luminance generated by scattering within the observer's eye).

Dropping the now unnecessary superindex s, and applying conditions (i)-(iv) to Eqs. (7)-(8) of Section 2 we have:

$$L_\beta(t) = N_\beta(t) \left\{ [1 - \gamma(t)] \mathcal{H}_{\beta,\text{HPS}} + \gamma(t) \mathcal{H}_{\beta,\text{LED}} \right\} \tag{9}$$

$$\frac{\dot{L}_\beta(t)}{L_\beta(t)} = \dot{n}_\beta(t) + \frac{\left(\frac{\mathcal{H}_{\beta,\text{LED}}}{\mathcal{H}_{\beta,\text{HPS}}} - 1\right) \dot{\gamma}}{1 + \left(\frac{\mathcal{H}_{\beta,\text{LED}}}{\mathcal{H}_{\beta,\text{HPS}}} - 1\right) \gamma} \tag{10}$$

where $\dot{n}_\beta(t) = \dot{N}_\beta(t)/N_\beta(t)$ is the relative rate of change of the total lumen emissions affecting the indicator β, expressed in yr$^{-1}$ or equivalently in %/yr.

Recall that the functions $N_\beta(t)$ (total amount of lumen emitted per unit area) and $\gamma(t)$ (fraction of the total $N_\beta(t)$ which is emitted by LED lamps) are in principle independent of each other. Their combinations may describe very general transformation processes. Equations (9)-(10) can be applied to particular cases in which the type of sources is changed without modifying the total emissions ($N_\beta(t)$ constant, $\dot{n}_\beta(t) = 0$), and to others in which the total emissions are changed while keeping constant the mix of sources ($\dot{\gamma}(t) = 0$). They can also be applied to the general case in which both functions vary, which is the subject of this section The same can be said of the previous Eqs. (1) to (8).

The approach to test the null hypothesis is based on estimating the rate of change of the emitted lumen $\dot{n}_\beta(t)$ using Eq. (10). The inputs for this calculation are the measured rates of change of the indicators $\dot{L}_\beta(t)/L_\beta(t)$, and the calculated values of the second term of the right-hand side of this equation, which accounts for the effects of the spectral shift. If the VIIRS-DNB and GAN measured radiances would respond to the emissions of the same set of sources, then one should find $\dot{n}_{\text{LED}}(t) = \dot{n}_{\text{GAN}}(t)$. Whereas the absolute amount of radiance and the effects of the spectral shift will generally be different for DNB and GAN, the rate of change of the emissions of the sources should be the same in both cases (if the null hypothesis holds).

*3.2 Evaluating the spectral shift term*

The reported DNB and GAN radiance rates are $\dot{L}_{\text{DNB}}(t)/L_{\text{DNB}}(t) = 2.2$ %/yr [26] and $\dot{L}_{\text{GAN}}(t)/L_{\text{GAN}}(t) = 9.6$ %/yr [25], respectively. In order to elucidade if these diverging rates may be explained by a combination of the spectral shift due to the change from HPS to LED (last term of Eq. (10)), and the evolution of the overall emissions affecting both indicators, $\dot{n}_\beta(t)$, the dimensionless numerical ratios $\mathcal{H}_{\text{DNB,LED}}/\mathcal{H}_{\text{DNB,HPS}}$ and $\mathcal{H}_{\text{GAN,LED}}/\mathcal{H}_{\text{GAN,HPS}}$ shall be determined. Recall that for an initial condition $\gamma(t_0) = 0$ at which all lamps are HPS, and a final condition $\gamma(t_1) = 1$ at which all lamps are LED, we have, from Eq. (9),

$$\frac{L_\beta(t_1)}{L_\beta(t_0)} = \frac{N_\beta(t_1)}{N_\beta(t_0)} \times \frac{\mathcal{H}_{\beta,\text{LED}}}{\mathcal{H}_{\beta,\text{HPS}}} \tag{11}$$

so that the $\mathcal{H}_{\beta,\text{LED}}/\mathcal{H}_{\beta,\text{HPS}}$ ratios may be obtained by calculating the ratios of the radiances at the start and the end of the transition period, $L_\beta(t_1)/L_\beta(t_0)$, using suitable theoretical models under the formal condition of constant lumen emissions $N(t_1) = N(t_0)$.

Some of these ratios are already available in the literature. The VIIRS-DNB $\mathcal{H}_{\beta k}$ terms for a standard configuration of the sources, atmosphere and observation geometry in a transition from a mix of HPS and metal halide lamps (with a residual contribution of mercury vapor sources) to 4000 K LED were calculated in [27]. They were found to be $\mathcal{H}_{\text{DNB,LED}} = 0.081 \times 10^{-3}$ and $\mathcal{H}_{\text{DNB,HPS}} = 0.107 \times 10^{-3}$, both in W·m$^{-2}$·sr$^{-1}$/(lm·m$^{-2}$). This suggests for the VIIRS-DNB observations a ratio $\mathcal{H}_{\text{DNB,LED}}/\mathcal{H}_{\text{DNB,HPS}} \approx 0.76$. For the present study we have recalculated these values for the transition from a uniform set of HPS lamps to 4000 K and 3000 K LED lamps (separately). The HPS and 4000 K lamp spectra were those used in [24], and the 3000 K spectrum was the LED_3045K of the lamp spectra database of the Laboratory of Scientific Instrumentation of the Universidad Complutense de Madrid (LICA-UCM) [28] (https://guaix.fis.ucm.es/lamps_spectra). The corresponding values are $\mathcal{H}_{\text{DNB,HPS}} = 0.138 \times 10^{-3}$, $\mathcal{H}_{\text{DNB,LED4K}} = 0.086 \times 10^{-3}$, and $\mathcal{H}_{\text{DNB,LED3K}} = 0.096 \times 10^{-3}$, all in W·m$^{-2}$·sr$^{-1}$/(lm·m$^{-2}$), from which $\mathcal{H}_{\text{DNB,LED4K}}/\mathcal{H}_{\text{DNB,HPS}} \approx 0.62$ and $\mathcal{H}_{\text{DNB,LED3K}}/\mathcal{H}_{\text{DNB,HPS}} \approx 0.70$.

For visual GAN observations, the ratio $\mathcal{H}_{\text{GAN,LED}}/\mathcal{H}_{\text{GAN,HPS}}$ arises from two contributions, both related to the scattering of artificial light. Recall that GAN radiances are estimated from unaided-eye observations of the artificial night sky brightness, using as a proxy the number of visible stars. This number is limited by the scattered artificial light reaching the retina, which reduces the luminance contrast of the stars against the night sky background. As the sky luminance increases, the dimmest stars progressively fall below the luminance contrast threshold of the observer and the number of visible stars decreases. Two main scattering processes have to be taken into account: the scattering of light in the terrestrial atmosphere (ATM) and the scattering within the observer's eye (intraocular scattering, IOC) [29]. Both contribute to the build-up of the anthropogenic luminance perceived by the sky observers in light polluted places.

From Eq. (5) we have

$$\frac{\mathcal{H}_{\text{GAN,LED}}}{\mathcal{H}_{\text{GAN,HPS}}} = \frac{\mathcal{H}_{\text{GAN,LED}}^{\text{ATM}} + \mathcal{H}_{\text{GAN,LED}}^{\text{IOC}}}{\mathcal{H}_{\text{GAN,HPS}}^{\text{ATM}} + \mathcal{H}_{\text{GAN,HPS}}^{\text{IOC}}} \qquad (12)$$

The relative contribution of the ATM and IOC scattering terms to the sky luminance perceived by the observers depends on multiple factors, including the spatial distribution of the light sources, their angular emission patterns and spectra, the reflective properties of the surrounding surfaces, the direction of the line of sight, the total lumen emissions and the adaptation state of the observer. At short distances from the luminaires (~10-20 m) the intraocular scattering radiance $L_{\text{GAN}}^{\text{IOC}}(t)$ due to direct emissions of nearby lamps can be larger than or equal to the zenith radiance $L_{\text{GAN}}^{\text{ATM}}(t)$ due to the atmospheric scattering of the ensemble of urban emissions, while at medium to long distances from the nearest luminaire the atmospheric scattering is the leading term [29]. Eq. (12) can be rewritten in a more convenient form, in terms of the ratio of intraocular and atmospheric scattered radiances at the initial time, $c \equiv L_{\text{GAN}}^{\text{IOC}}(t_0)/L_{\text{GAN}}^{\text{ATM}}(t_0)$. From our assumed initial conditions ($\gamma(t_0) = 0$, all lamps HPS), it follows $c = \mathcal{H}_{\text{GAN,HPS}}^{\text{IOC}}/\mathcal{H}_{\text{GAN,HPS}}^{\text{ATM}}$ and Eq. (12) has the equivalent form:

$$\frac{\mathcal{H}_{\text{GAN,LED}}}{\mathcal{H}_{\text{GAN,HPS}}} = \frac{1}{(1+c)} \left( \frac{\mathcal{H}_{\text{GAN,LED}}^{\text{ATM}}}{\mathcal{H}_{\text{GAN,HPS}}^{\text{ATM}}} + c \frac{\mathcal{H}_{\text{GAN,LED}}^{\text{IOC}}}{\mathcal{H}_{\text{GAN,HPS}}^{\text{IOC}}} \right) \qquad (13)$$

Equation (13) allows treating separately the effects of the atmospheric and the intraocular processes by using their own ratios $\mathcal{H}_{\text{GAN,LED}}^{\text{ATM}}/\mathcal{H}_{\text{GAN,HPS}}^{\text{ATM}}$ and $\mathcal{H}_{\text{GAN,LED}}^{\text{IOC}}/\mathcal{H}_{\text{GAN,HPS}}^{\text{IOC}}$.

The parameter space of this problem is exceedingly large and it seems not feasible to analyze all possible configurations relevant for GAN observations. However, some estimations can be made for reasonable conditions of observation.

The dependence of the skyglow on the light source spectra for different types of lamp technologies in the photopic and scotopic adaptation regimes has been thoroughly analyzed by Luginbuhl et al [30]. The evolution of the artificial zenith sky radiance $L_{\text{GAN}}^{\text{ATM}}(t)$ for the transition HPS to 4000 K LED has been calculated in [24]. Recall that the ratio $\mathcal{H}_{\text{GAN,LED}}^{\text{ATM}}/\mathcal{H}_{\text{GAN,HPS}}^{\text{ATM}}$ is equal to the value of $L_{\text{GAN}}^{\text{ATM}}(t_1)/L_{\text{GAN}}^{\text{ATM}}(t_0)$ for constant lumen emissions ($N(t_1)/N(t_0) = 1$) and $\mathcal{H}_{\text{GAN},k}^{\text{ATM}}$ terms independent of time. This situation corresponds to the top-left panel of Figure 6 of [24]. The $L_{\text{GAN}}^{\text{ATM}}(t_1)/L_{\text{GAN}}^{\text{ATM}}(t_0)$ photopic value for a distance source-observer 0.1 km, and the scotopic ones for several distances can be deduced from the value of the curves in that Figure for the abscissa equal to 1 (all lamps LED, $\gamma = 1$). The normalized radiances $L_{\text{GAN}}^{\text{ATM}}(t)/L_{\text{GAN}}^{\text{ATM}}(t_0)$ are expressed in the vertical axis of that plot as astronomical magnitude changes $\Delta m(t)$ with respect to the initial conditions, so that $L_{\text{GAN}}^{\text{ATM}}(t)/L_{\text{GAN}}^{\text{ATM}}(t_0) = 10^{-0.4 \times \Delta m(t)}$.

For the present work we have computed the values of $\mathcal{H}_{\text{GAN,LED}}^{\text{ATM}}/\mathcal{H}_{\text{GAN,HPS}}^{\text{ATM}}$ for several distances source-observer in the transitions HPS to 4000 K and 3000 K LEDs. The calculations were made for the photopic, mesopic (with adaptation coefficient 0.5) [31], and scotopic spectral sensitivity bands The atmospheric conditions and the ground reflectance were set as in ref. [27], and the angular radiant intensity emission function was the one used by Falchi et al. in [32]. The Lambertian and the intermediate-angle terms of this function were applied to the light reflected on the pavements, whereas the low-angle term was applied to the direct lamp spectra. The results are summarized in Table 1 below.

**Table 1.** Values of $\mathcal{H}_{\text{GAN,LED}}^{\text{ATM}}/\mathcal{H}_{\text{GAN,HPS}}^{\text{ATM}}$ for the atmospheric scattering term

| Distance source-observer (km) | $\mathcal{H}_{\text{GAN,LED}}^{\text{ATM}}/\mathcal{H}_{\text{GAN,HPS}}^{\text{ATM}}$ (photopic) | $\mathcal{H}_{\text{GAN,LED}}^{\text{ATM}}/\mathcal{H}_{\text{GAN,HPS}}^{\text{ATM}}$ (mesopic) | $\mathcal{H}_{\text{GAN,LED}}^{\text{ATM}}/\mathcal{H}_{\text{GAN,HPS}}^{\text{ATM}}$ (scotopic) |
|---|---|---|---|
| 4000 K | | | |
| 0.1 | 1.063 | 1.472 | 2.813 |
| 0.5 | 1.061 | 1.469 | 2.808 |
| 1.0 | 1.060 | 1.464 | 2.802 |
| 5.0 | 1.048 | 1.435 | 2.763 |
| 10 | 1.037 | 1.407 | 2.725 |
| 50 | 1.014 | 1.348 | 2.634 |
| 100 | 0.989 | 1.289 | 2.539 |
| *Mean value* | *1.039* | *1.412* | *2.727* |
| 3000 K | | | |
| 0.1    | 1.040 | 1.287 | 2.095 |
| 0.5    | 1.039 | 1.285 | 2.092 |
| 1.0    | 1.038 | 1.282 | 2.088 |
| 5.0    | 1.031 | 1.263 | 2.061 |
| 10     | 1.023 | 1.246 | 2.037 |
| 50     | 1.009 | 1.209 | 1.979 |
| 100    | 0.992 | 1.173 | 1.924 |
| *Mean value* | *1.025* | *1.249* | *2.039* |

**Table 2.** Values of $\mathcal{H}_{\text{GAN,LED}}^{\text{IOC}}/\mathcal{H}_{\text{GAN,HPS}}^{\text{IOC}}$ for the intraocular scattering term [see Appendix]

| Observer group | $\mathcal{H}_{\text{GAN,LED}}^{\text{IOC}}/\mathcal{H}_{\text{GAN,HPS}}^{\text{IOC}}$ (photopic) | $\mathcal{H}_{\text{GAN,LED}}^{\text{IOC}}/\mathcal{H}_{\text{GAN,HPS}}^{\text{IOC}}$ (mesopic) | $\mathcal{H}_{\text{GAN,LED}}^{\text{IOC}}/\mathcal{H}_{\text{GAN,HPS}}^{\text{IOC}}$ (scotopic) |
|---|---|---|---|
| 4000 K | | | |
| > 40 yr | 1.031 | 1.407 | 2.799 |
| Blue-eyed | 1.008 | 1.348 | 2.713 |
| Green and light brown-eyed | 1.013 | 1.373 | 2.758 |
| Brown-eyed | 1.097 | 1.503 | 2.915 |
| Dark brown-eyed | 1.133 | 1.595 | 2.965 |
| *Mean value* | *1.056* | *1.445* | *2.830* |
| 3000 K | | | |
| > 40 yr | 1.035 | 1.264 | 2.113 |
| Blue-eyed | 1.023 | 1.225 | 2.036 |
| Green and light brown-eyed | 1.027 | 1.237 | 2.046 |
| Brown-eyed | 1.093 | 1.333 | 2.169 |
| Dark brown-eyed | 1.097 | 1.376 | 2.204 |
| *Mean value* | *1.055* | *1.287* | *2.114* |

The remaining ratios to be calculated are those corresponding to the intraocular scattering. Its equivalent radiance, $L_{\text{GAN}}^{\text{IOC}}(t)$, can be described by means of the glare point spread function [28,32,33]. A general formulation based on the spectral straylight parameter $s(\lambda)$ is developed in the Appendix. The strength and the spectral behavior of the intraocular scattering depend on the demographics of the observers [33,34]. The spectral $s(\lambda)$ functions used in this work are based on the discrete sets of narrowband measurements reported by Coppens et al. [35] for different populations, at wavelengths 457, 503, 548, 583 and 625 nm. For the purpose of the present calculations, the Coppens et al. values for each observer group were linearly interpolated within the data range, and linearly extrapolated to span the remaining part of the visible regions of the spectrum below 457 nm and above 625 nm. The Coppens et al. populations are defined by age (>40 yr old) or eye color and pigmentation (blue, green and light-brown, brown, and dark-brown eyes). The resulting $\mathcal{H}_{\text{GAN,LED}}^{\text{IOC}}/\mathcal{H}_{\text{GAN,HPS}}^{\text{IOC}}$ ratios for each observer group in the photopic, mesopic and scotopic bands, for the transitions HPS to 4000 K and 3000 K LEDs are listed in Table 2. These ratios are consistently larger than 1, and larger for scotopic than for photopic observers, reflecting the fact that high CCT LEDs give rise to more intense glare effects due to their high blue component. The smallest values of these ratios correspond to weakly pigmented eyes (blue eyes), for which the short wavelength scattering of the iris, cornea and eye lens is partially balanced by the long-wavelength scattering from the highly vascularized retina [33,34]. The values displayed in Table 2 correspond to the dominant intraocular scattering process, that is, the scattering of the lamps' light directly entering the eyes. Note that the light scattered by the atmosphere (ATM) is also partially scattered again within the eye. This is a second-order effect which was not included here. In cases where it could be deemed relevant, it could be accounted for according to Eq. (5) of [29].

For the numerical calculations in this section we used the mean values (column-wise) of the ratios listed in tables 1 and 2. Each observer adaptation state (photopic, mesopic, and scotopic), and LED CCT was studied separately. The mean ATM and IOC ratios were combined into a single ratio for each state according to Eq. (13) with $c = 1$ (equal contribution of both processes under HPS illumination), resulting in $\mathcal{H}_{\text{GAN,LED}}/\mathcal{H}_{\text{GAN,HPS}} =$[ 1.048  1.429  2.778] for the transition HPS to 4000 K LED under photopic, mesopic, and scotopic adaptation, respectively. It is not precisely known whether GAN observers do intentionally avoid making the observations in the neighborhood of streetlights and other glaring sources, or may try to avoid glare in some other way, for example screening the lights with their hands, thus giving more relative weight to the ATM contribution. As shown in [29] the ATM and IOC luminances (without screening) would be approximately equal at distances ~15-20 m from the light pole. The assumption of equal contribution is only an approximate choice made for the purpose of this calculation. Note however that, for the same adaptation state, the mean ATM and IOC ratios shown in the last row of tables 1 and 2 are very close, essentially because both reflect the wavelength dependence of similar scattering phenomena. For arbitrary values of $c$, the actual ratio $\mathcal{H}_{\text{GAN,LED}}/\mathcal{H}_{\text{GAN,HPS}}$ shall be comprised in the interval defined by the extreme situations $c=0$ (only atmospheric scattering, no intraocular, mean value row of table 1), and $c \to \infty$ (only intraocular scattering, no atmospheric, mean value row of table 2), that is, within 1.039-1.056 (photopic), 1.412-1.445 (mesopic with adaptation coefficient 0.5), and 2.727-2.830 (scotopic). In consequence, other choices for $c$ would not dramatically alter the results. A similar argument can be made for the transition to 3000 K LED, whose ratios $\mathcal{H}_{\text{GAN,LED}}/\mathcal{H}_{\text{GAN,HPS}}$ for $c = 1$ are [1.040, 1.268, 2.077] for photopic, mesopic, and scotopic observers, respectively.

As a last step to apply Eq. (10) to our present problem, it is necessary to estimate the form of the function $\gamma(t)$ and its time derivative, $\dot{\gamma}(t)$. Recall that in a two-source type transition process $\gamma(t)$ is the ratio of the lumen per unit area emitted by the new sources (in our case, LED) to the total ones (LED+HPS). Its time derivative is not precisely known, due to the lack of systematic lamp inventory updates, but it is expected to vary widely from one region of the world to another. To get some insight about the order of magnitude of its effects, we assume here a linear evolution at a constant time rate along a period of $T_0$ years, such that $\gamma(t) = t/T_0$ and $\dot{\gamma}(t) = 1/T_0$ (expressed in 1/yr or %/yr). The calculations in this section are made for three transition scenarios, $T_0 =$10, 20, and 30 years, corresponding to substitution rates $\dot{\gamma}(t)=$ 10 %/yr, 5 %/yr, and (10/3) %/yr, respectively.

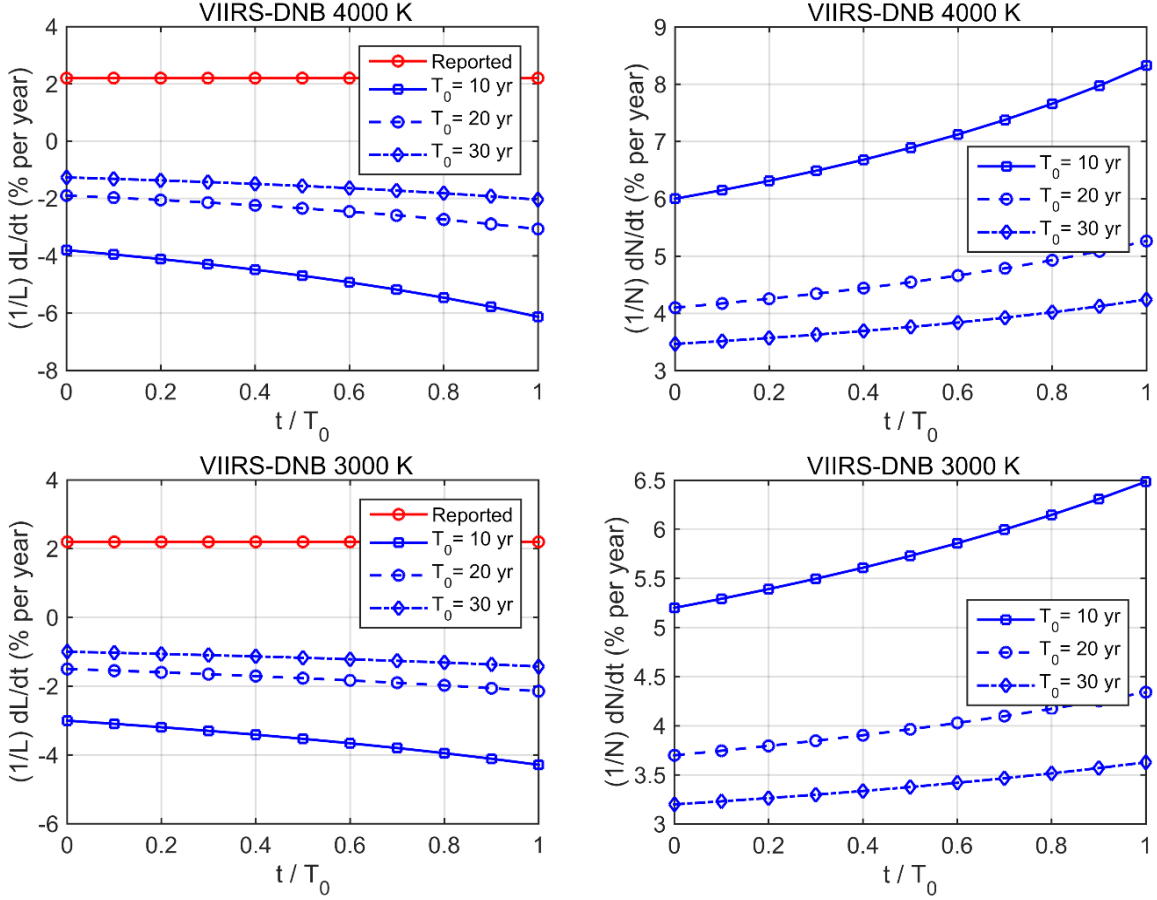

**Fig. 1.** *Left panel:* Radiance change rate $\dot{L}_{DNB}(t)/L_{DNB}(t)$, in % per year, vs. normalized time $t/T_0$ for the VIIRS-DNB. The constant line labeled 'Reported' corresponds to the published rate 2.2 %/yr. Lines labeled with $T_0$ (in years) display the last term of the right-hand side of Eq. (10), that is, the radiance change rate associated with the spectral shift HPS to LED at constant lumen emissions, calculated for $\mathcal{H}_{DNB,LED}/\mathcal{H}_{DNB,HPS} = 0.76$. *Right panel:* estimated rates of increase of $\dot{n}_{DNB}(t) = (1/N_{DNB}(t))\,dN_{DNB}(t)/dt$, the relative change of the lumen emissions per unit area of the sources sensed by the VIIRS-DNB, in %/yr. They are calculated according to Eq. (10) as the difference between the measured value and the spectral shift terms shown in the left panel. The upper row corresponds to the transition HPS to 4000 K LED, the lower row to HPS to 3000 K LED.

Figure 1 displays the results for the VIIRS-DNB radiance. The upper row corresponds to the transition HPS to 4000 K LED, the lower row to HPS to 3000 K LED. The vertical axis of the left panels shows the rate of change $\dot{L}_{DNB}(t)/L_{DNB}(t)$, expressed in % per year, and the horizontal axis the normalized time, $t/T_0$. The constant line labeled 'Reported' corresponds to the published value $\dot{L}_{DNB}(t)/L_{DNB}(t) = 2.2$ %/yr [26]. The lines labeled with periods $T_0$ in years correspond to the values of the last term of the right-hand side of Eq. (10), that is, the expected change rate of the observed radiance due to the spectral shift of the sources, at constant lumen emissions, for the three scenarios here considered. As expected and previously reported [24,27], the VIIRS-DNB radiance rates associated with the spectral change from HPS to 4000 K or 3000 K LED at constant emitted lumen are negative. This is due, among other things, to the progressive reduction of the contribution of the HPS 819 nm near-infrared line and the inability of the DNB band to detect the emissions of the 450 nm blue peak of phosphor-coated LED. From Eq. (10) it also follows that the difference between the reported rate and these spectral terms is $\dot{n}_{DNB}(t) = [1/N_{DNB}(t)]\, dN_{DNB}(t)/dt$, the relative rate of change of the total emissions in lumen per unit area of the light sources sensed by the VIIRS-DNB. The values of $\dot{n}_{DNB}(t)$, expressed in %/yr, are displayed in the right panels of Fig.1. The rate of increase of the emissions of the sources sensed by the VIIRS-DNB corrected from the spectral shift is then expected to lie within the range 3%/yr-6%/yr for 4000 K and 3.2%/year-5.2%/year for 3000 K at the start of the transformation process, depending on the celerity of this process, $1/T_0$. These values monotonically increase with the time of observation, $t$.

The results for GAN observations are summarized in Fig. 2 and 3, for 4000 K and 3000 K, respectively. The panels in the left column display the evolution of the radiance rates $\dot{L}_{GAN}(t)/L_{GAN}(t)$, expressed in % per year, for observers with photopic, mesopic and scotopic adaptation (from top to bottom), resulting from the combination of the atmospheric and intraocular scattering. The 'Reported' line corresponds to the published value $\dot{L}_{GAN}(t)/L_{GAN}(t) = 9.6 \pm 0.4$ %/yr [25]. The lines labeled with different periods $T_0$ in years correspond to the spectral term of the right-hand side of Eq. (10). The panels in the right column display the resulting $\dot{n}_{GAN}(t)$, the estimated rate of change of the emissions of the light sources that are sensed by GAN observers. The lines for the VIIRS-DNB (same $\dot{n}_{DNB}(t)$ as in Fig 1, right column) have been added to these panels for reference (thin red lines). If the set of sources sensed by GAN observers were the same as the ones sensed by the VIIRS-DNB, then we should have $\dot{n}_{GAN}(t) = \dot{n}_{DNB}(t)$ and the corresponding curves would be coincident for all values of $t$. The inspection of the right column panels of these figures clearly reveals that this equality for all $t$ is not fulfilled for any observer adaptation state, under the transition conditions assumed in this section (HPS to LED lumen replacement at constant rate). If these assumptions are correct, the results strongly suggest the existence of sources affecting GAN that do not show up in VIIRS-DNB measurements (and vice-versa). The emissions of both sets of sources may increase with the same rate at some particular times and under certain transition conditions, when the corresponding curves $\dot{n}_{GAN}(t)$ and $\dot{n}_{DNB}(t)$ interesect. This may happen for scotopic observers (and probably mesopic ones, for adaptation coefficients smaller than 0.5). As an example, equal VIIRS-DNB and GAN emissions increase rates could happen for scotopic observers in a 4000 K transition scenario of $T_0 = 20$ years at $t/T_0 = 0.4$, for which $\dot{n}_{GAN}(t) \approx \dot{n}_{DNB}(t) \approx 4.4$%/yr (lower right panel of Fig. 2).

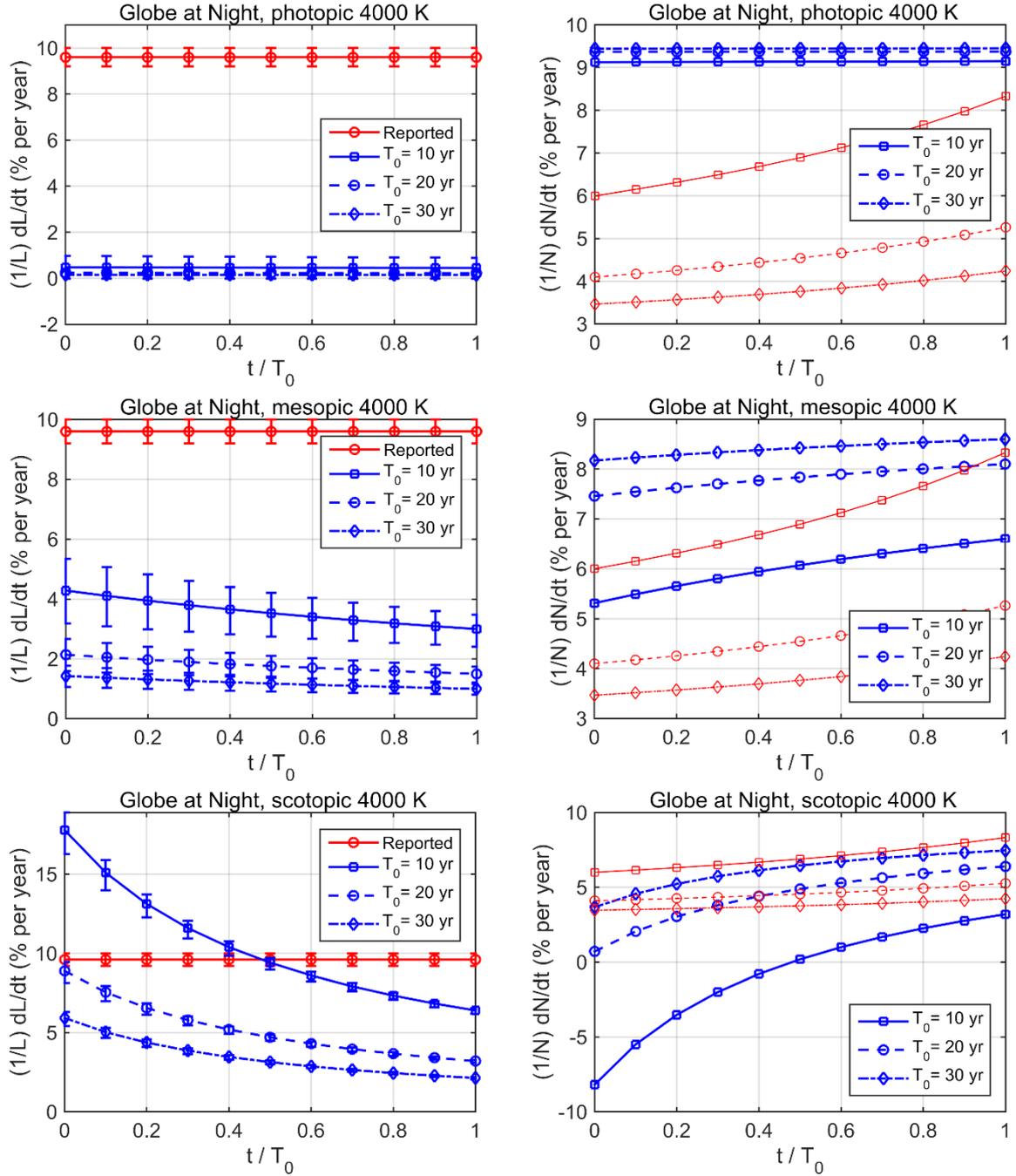

**Fig. 2.** GAN rates for a HPS to 4000 K LED transition. *Left column:* Radiance change rate $\dot{L}_{\mathrm{GAN}}(t)/L_{\mathrm{GAN}}(t)$, in % per year, vs. normalized time $t/T_0$ for (*top to bottom*) photopic, mesopic, and scotopic observers. The constant line labeled 'Reported' corresponds to the published rate $9.6 \pm 0.4$ %/$yr$. Lines labeled with $T_0$ (in years) display the last term of the right-hand side of Eq. (10), that is, the radiance change rate associated with the spectral shift HPS to LED at constant lumen emissions, calculated for the mean values of $\mathcal{H}_{\mathrm{GAN,LED}}^{\mathrm{ATM}}/\mathcal{H}_{\mathrm{GAN,HPS}}^{\mathrm{ATM}}$ and $\mathcal{H}_{\mathrm{GAN,LED}}^{\mathrm{ATM}}/\mathcal{H}_{\mathrm{GAN,HPS}}^{\mathrm{ATM}}$ for each observer adaptation state listed in Tables 1 and 2, combined according to Eq. (13) with $c = 1$. *Right column:* estimated rates of increase of $\dot{n}_{\mathrm{GAN}}(t) = [1/N_{\mathrm{GAN}}(t)]\,\mathrm{d}N_{\mathrm{GAN}}(t)/\mathrm{d}t$, the relative change of the lumen emissions per unit area of the sources sensed by the GAN observers, in %/yr. They are calculated from Eq. (10) as the difference between the reported value and the spectral

shift terms. The curves for the VIIRS-DNB, $\dot{n}_{DNB}(t)$, are displayed as thin red lines for comparison.

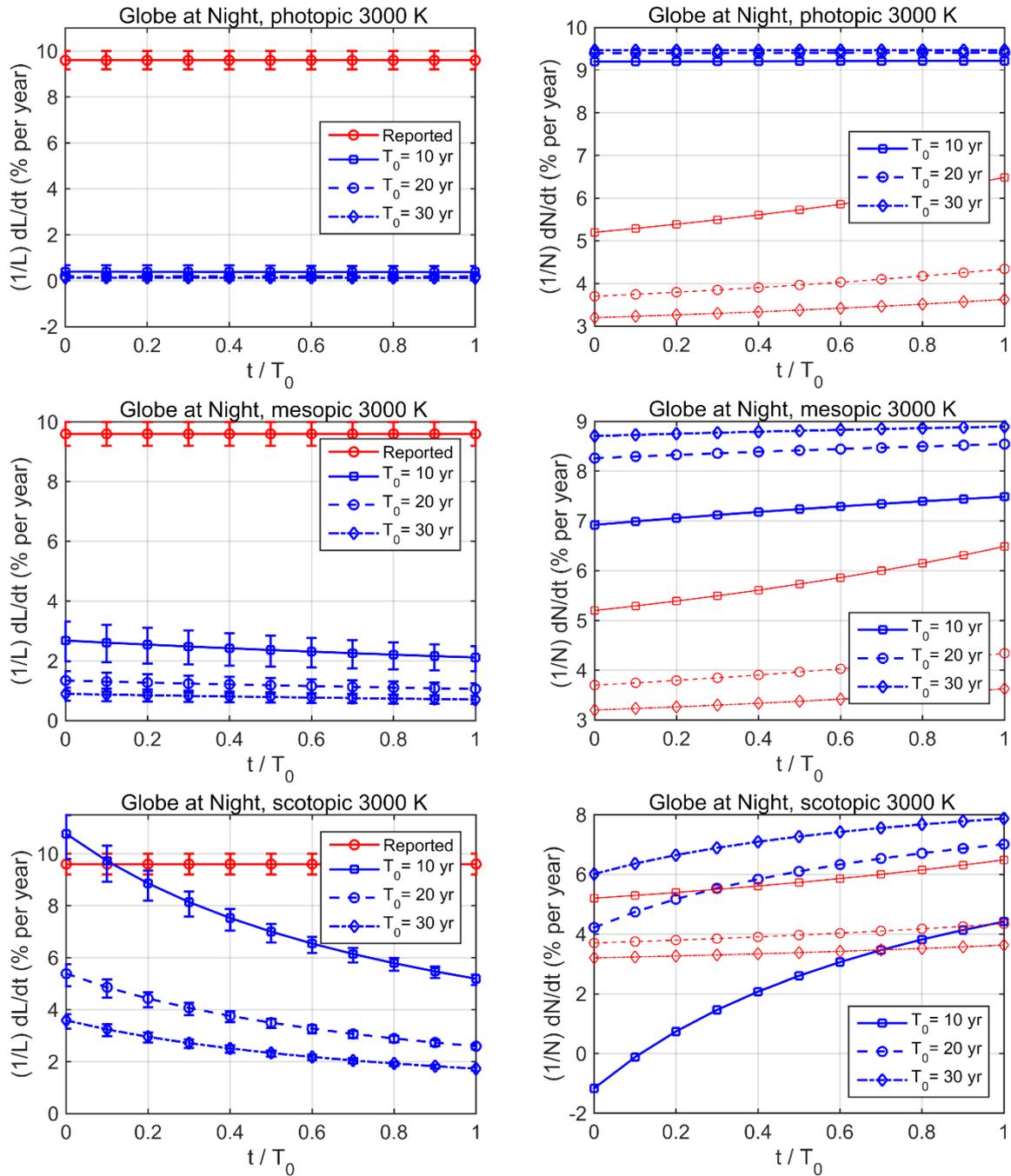

**Fig. 3.** GAN rates for a HPS to 3000 K LED transition. See caption of Fig. 2 for details.

## 4. Discussion

The results in Section 3 provide some partial answers to the question that motivates this paper. The main outcome is that the mere change of spectrum in the transition HPS to 4000 K or 3000

K LED does not seem to be enough to explain the diverging evolution rates of VIIRS-DNB and GAN radiances. If the VIIRS-DNB and GAN observers were sensing the effects of the same ensemble of light sources, the rates of increase of their lumen emissions deduced from Eq. (10) should coincide along the whole transition process, $\dot{n}_{GAN}(t) = \dot{n}_{DNB}(t)$ ($0 \leq t \leq T_0$). However, the differences between these rates are apparent (Fig. 2 and 3, right column), excepting for very particular conditions. This opens the way to considering the existence of specific sets of light sources affecting in a different way each type of measurements.

The differences between $\dot{n}_{GAN}(t)$ and $\dot{n}_{DNB}(t)$ are larger for photopic observers. After correcting the GAN observations for the spectral shifts of the HPS to LED transition, the overall emissions affecting photopic GAN observers are expected to grow at a rate ~9% per year, substantially faster than the emissions detected by VIIRS-DNB, which for $T_0 = 20$ yr should actually grow ~4.1-5.2%/yr for 4000 K and ~3.7-4.3%/yr for 3000 K, depending on $t$, after spectral correction (Fig. 1, right column). Of course, it shall not be excluded the possibility that GAN observers also sense in some proportion the lights that are detected by the VIIRS-DNB, in addition to the GAN-specific ones. In this case, however, the latter should grow faster than the observed value $\dot{L}_{GAN}(t)/L_{GAN}(t)$=9%/yr, in order to compensate for the slower rate of the former. For instance, if both sets of lights would contribute the same amount of radiance to the GAN observations at a given time, the GAN-specific light emisions (in lumen per unit area) should grow at a rate ~13-14%/yr.

The differences in the estimated emission rates are smaller for mesopic and scotopic GAN observers, since the spectral shifts have stronger effects in these visual bands. It is even possible to attain equal expected rates, $\dot{n}_{GAN}(t) = \dot{n}_{DNB}(t)$, for some particular combinations of $t$ and $T_0$ (see e.g. the intersection between corresponding lines for $T_0 = 20$ yr in the scotopic panel, right column of Fig. (2)). It is debatable whether scotopic adaptation could be acting in a sufficiently large subset of GAN observations, due to the prevailing high illumination levels in populated areas of our planet. Different levels of mesopic adaptation, in turn, are definitely possible.

If there are no significant biases in the reported rates $\dot{L}_{GAN}(t)/L_{GAN}(t)$ and $\dot{L}_{DNB}(t)/L_{DNB}(t)$ [25,26], the results of Section 3 strongly suggest the existence of GAN-specific light sources. This additional set of emissions, affecting the GAN observations but not the VIIRS-DNB ones, could be attributed to several processes. An obvious candidate would be an increase in the amount of ornamental, commercial, vehicle and advertising lights, including high-luminance LED billboards located in the immediate vicinity of the average GAN observer, which are strongly detrimental for star watching. These sources are usually switched on during the first half of the night, a typical time period for unaided-eye GAN observations, and switched off or considerably dimmed in many places at midnight, much earlier than the usual time of VIIRS-DNB data acquisition (~01:30 local solar time). These light sources, especially if blocked by nearby obstacles, could make a small contribution to the total urban light emissions in near-zenith directions within the pixels of mid-resolution satellites (a few hundred meter wide), but could increase substantially the scattered radiance in the first meters of the local air column [36], and also inside the observer eye if the lamps are directly visible [29], leading to progressively larger equivalent veiling luminances and corresponding loss of star visibility. Intraocular scattering has an important influence on the perception of sky brightness and star detection in observations made close to light sources (such as street lamps) [29], since it is widely known that as intraocular scattering increases, certain visual functions such as contrast sensitivity or visual discrimination ability are impaired [37]. In addition, intraocular scattering is wavelength dependent, so that for several observer populations there is a greater contribution of short wavelengths as other authors have shown [35]. The LED sources that are replacing high-pressure sodium lamps have a significant spectral component of short wavelengths, which would contribute to a greater amount of intraocular scattering in these populations. In fact, it has been clinically demonstrated with young observers that straylight increases for short wavelengths and the visual function under low-illumination conditions worsens [38].

Note that what matters here is not so much the absolute amount of these GAN-specific emissions, but their yearly growth rate. As far as we know, there are no published reports on the rate of increase of the emissions from these types of ground sources to which the VIIRS-DNB is almost blind. An average rate equal to or larger than ~9 %/yr worldwide might seem excessive, based on anecdotal evidence. However, this topic deserves to be further explored. Citizen science campaigns [39] and comprehensive analyses of administrative authorizations for ornamental, commercial and advertising lighting installations could help to settle this relevant issue.

The formulation in section 2 of this paper is completely general and can be applied to a large class of problems related to the time evolution of light pollution indicators. The results in section 3, in turn, should be taken with caution, due to the need of making several assumptions to cope with the lack of information.

One of these assumptions concerns the rate of replacement of technologies, $\gamma(t)$. The values of $\dot{n}_{\mathrm{DNB}}(t)$ and $\dot{n}_{\mathrm{GAN}}(t)$ estimated in section 3 are dependent on the assumed form of $\gamma(t)$ and $\dot{\gamma}(t)$. The latter functions, which inform of the fraction of LED lumen emissions relative to the total HPS+LED at time $t$, and of its change in %/year, respectively, are very likely highly variable from one world region to another, and reliable statistical data do not appear to be widely available. Reports from individual countries provide some insights about these differences, which arguably also exist at lower territorial aggregation levels. As of 2020, the LED sources represented a 55% of the total streetlights in the UK [40], being the remaining ones HPS and, in smaller quantities, low-pressure sodium, mercury vapor, and other high-intensity discharge lamps. At that time the HPS to LED conversion was progressing at a rate of 3 %/yr. A US report published the same year [41] quoted a cumulative installed fraction of LED lamps in outdoor lighting of 27.2% in 2016 and 51.4% in 2018. Whereas the $\gamma(t)$ values for 2018 (US) and 2020 (UK) are of the same order of magnitude, the substitution rate per year seems to be substantially larger in the US case. Note however that these published data refer to the fractions of lamps, not to the fractions of lumen emitted by each type of technology, the latter being what $\gamma(t)$ stands for.

Another simplification made in section 3 is the assumption of the constancy of the $\mathcal{H}_{\beta k}^{s}(t)$ factors, whereby $\dot{\mathcal{H}}_{\beta k}^{s}(t) = 0$. Recall that $\mathcal{H}_{\beta k}^{s}(t)$ is the radiance detected in the β-band per lm·m$^{-2}$ emitted by the $k$-th type of lamps of the $s$-th set of light sources. The term with its derivative in Eq. (8) accounts for the changes in the detected radiance that are not due to spectral shifts nor to the variation in the number of emitted lm·m$^{-2}$. This factor may evolve in time if e.g. the angular emission pattern of the (*s,k*) subset of light sources does change. The main driver of these changes could probably be the introduction of ultra-bright vertical LED billobards with angular emission functions peaking at near horizontal directions. This would expectedly lead to an increase of the $\mathcal{H}_{\text{GAN,LED}}$ factor, due to the relative increase of the light scattered from the nearby atmosphere and the glaring effects of these billboards, stronger in comparison with traditional streetlights. The values of $\mathcal{H}_{\text{DNB,LED}}$ would very likely be much less affected, since the direct light from the billboards only reachs the satellite at large nadir angles from unobstructed locations, and the scattered light detected from orbit usually makes a relatively small fraction of the total radiance of the urban pixels. The net effect of the increase in $\mathcal{H}_{\text{GAN,LED}}$ would be an estimate of the rate of increase of the GAN-specific emissions smaller than the one quoted in Section 3.

As a matter of fact, the lack of precise information about some key parameters makes that the results in Section 3 should be considered as a first approximation to this problem, and as an example of application of the general method described in section 2. The key issue underlying the question that motivated this paper is whether the VIIRS-DNB data are a good proxy to estimate the actual evolution of the light emissions affecting observers on ground. This is not only relevant for assessing the degradation of the visual appearance of the night sky, but also for ecological and public health studies, as well as for verifying compliance with light emission control policies. Settling this interesting and still open issue requires collecting and analyzing finer-grained data on the actual ground observation conditions, including the luminance adaptation state and demographics of the observers, as well as the spectral shift rates at local sites. Besides, information about the state of the atmosphere at the times of observation should also be collected, to reduce the uncertainty derived from the variability of the atmospheric conditions [42].

## 5. Conclusions

We describe in this paper a general model for evaluating the time evolution of different light pollution indicators. It allows to disentangle the effects of actual changes in the total amount of light emissions from the effects of the spectral shifts in the emitted light. The model was applied here to analyze the remarkable difference between the reported rates of increase of the radiance measured by the VIIRS-DNB instrument on board the Suomi-NPP satellite, which averages to 2.2 %/yr worldwide, and that deduced from the Globe at Night unaided-eye star observations (GAN), which averages to 9.6 %/yr globally.

If both series of observations (GAN and VIIRS-DNB) would respond to the same set of light sources, the changes in their lumen per unit area relative emissions in percent per year, after correcting for spectral shift, should be the same for both indicators along the whole measurement period. However, the available evidence does not tend to support this assumption. If the VIIRS-DNB and GAN measurements are not biased, this strongly suggests the existence of different sets of light sources, evolving at different paces, affecting in different way the VIIRS-DNB and GAN observations. This result is relevant for assessing whether the VIIRS-DNB data are a good proxy for estimating the actual evolution of the light emissions experienced by ground observers.

In this work we assumed an outdoor lights transition from HPS to 4000 K or 3000 K LED with different time span scenarios, from $T_0$ =10 to $T_0$ =30 years. For the particular parameters used in this calculation and $T_0$ =20 yr, the actual rate of increase of the emissions of the sources sensed by the VIIRS-DNB after correcting the measurements for spectral shifts is estimated to be in the range ~4.1-5.2%/yr for 4000 K and ~3.7-4.3%/yr for 3000 K, depending on the year of observation. The emissions in lumen per unit area affecting photopic GAN observers, however, are estimated to increase at an actual rate of at least ~9 %/yr. The differences are smaller for mesopic and scotopic observers, for whom GAN light emission rates compatible with the VIIRS-DNB ones could be attained for appropriate combinations of the adaptation state, observation year, and transition scenario, without the need of resorting to GAN-specific light sources. Notwithstanding that, this compatibility only holds for very particular adaptation conditions and moments in time.

Settling the issue of whether different ensembles of light sources actually growing at substantially different rates are sensed by the VIIRS-DNB and GAN observers and/or whether there may be some bias in the reported rates of increase of the measured VIIRS-DNB and estimated GAN radiances requires gathering more detailed information. In particular, reliable data on the progress of the lamp substitution processes in different regions of the world, about the state of the atmosphere at the times of observation, and on the observer adaptation state and ocular media conditions.

**APPENDIX: Spectrally-resolved glare model**

The equivalent glare luminance $L_{eq}(\theta)$ in cd·m$^{-2}$ produced by intraocular scattering is commonly related to the glaring illuminance $E_{gl}$ on the eye pupil, in lx, through the glare point spread function $\Psi(\theta)$, with units sr$^{-1}$, defined as [33,34]:

$$L_{eq}(\theta) = \Psi(\theta)\, E_{gl} \tag{A1}$$

where $\theta$ is the angle between the observer line of sight and the position of the pointlike glaring source in the visual field, usually expressed in degrees (deg).

Several expressions for $\Psi(\theta)$ can be found in the literature, with increasing levels of complexity as they aim to describe with progressively better accuracy the angular dependence, including factors like the age of the observers and the degree of pigmentation of their ocular structures [33,34]. For the purposes of this work we will use the simple inverse squared Stiles-Holladay model $\Psi(\theta) = s\,\theta^{-2}$, where $s$ is the straylight parameter, with dimensions deg²·sr⁻¹, which provides information about the strength of the scattering within the eye.

The parameter $s$ depends on the spectral composition of the pupil glaring illuminance and on the characteristics of the observers. The spectral dependence of $s$ has been studied by Coppens et al. [35], using narrow-band filters at five discrete wavelengths in the interval 457nm-625nm, and by Castro-Torres et al. [38] in the RGB channels.

Here we provide a general expression of the value of $s$ for light of arbitrary spectrum. To formalize it, let us denote by $E_p(\lambda)$ the spectral irradiance produced by the glare source on the eye pupil, in units W·m⁻²·nm⁻¹. The pupil glare illuminance due to the irradiance contained in the spectral interval $[\lambda, \lambda + d\lambda]$ is then $dE_{gl}(\lambda) = K_r\,V(\lambda)\,E_p(\lambda)\,d\lambda$, where $V(\lambda)$ is the photopic spectral sensitivity function of the human visual system [7], and $K_r = 683$ lm·W⁻¹. The corresponding equivalent glare luminance at an angle $\theta$ can be written as $dL_{eq}(\theta, \lambda) = K_r\,V(\lambda)\,L(\theta, \lambda)\,d\lambda$, where $L(\theta, \lambda)$ is the spectral glare radiance, in W·m⁻²·sr⁻¹·nm⁻¹. Both quantities can be related through a spectral glare PSF that in general will depend not only on $\theta$ but also on $\lambda$, $\Psi(\theta, \lambda)$. So we have

$$dL_{eq}(\theta, \lambda) = \Psi(\theta, \lambda)\,dE_{gl}(\lambda) \tag{A2}$$

from which

$$L(\theta, \lambda) = \Psi(\theta, \lambda)\,E_p(\lambda) \tag{A3}$$

which is analog to Eq. (A1), but now written in terms of the *spectral* radiances and irradiances, instead of band-integrated luminances and illuminances. The units of $\Psi(\theta, \lambda)$ are sr⁻¹. For most applications of interest it can be assumed that $\Psi(\theta, \lambda)$ is a separable function, such that

$$\Psi(\theta, \lambda) = s(\lambda)\,\psi(\theta) \tag{A4}$$

where $s(\lambda)$ is the spectral straylight factor, determined from measurements made with quasi-monochromatic light, and $\psi(\theta)$ is a function carrying the angular dependence of glare, e.g. $\psi(\theta) = \theta^{-2}$ for the simplest Stiles-Holladay model.

From the above results, expressions for the photopic and scotopic equivalent glare luminances can be easily derived. Recalling the definitions of the photopic $L_{eq}(\theta)$ and $E_{gl}$ we have

$$L_{eq}(\theta) = K_r \int V(\lambda)\,L(\theta, \lambda)\,d\lambda = K_r \int V(\lambda)\,\Psi(\theta, \lambda)\,E_p(\lambda)\,d\lambda \tag{A5}$$

$$E_{gl} = K_r \int V(\lambda)\,E_p(\lambda)\,d\lambda \tag{A6}$$

and, from (A1), (A5), (A6),

$$\Psi(\theta) = \frac{\int V(\lambda)\,\Psi(\theta,\lambda)\,E_{\mathrm{p}}(\lambda)\,\mathrm{d}\lambda}{\int V(\lambda)\,E_{\mathrm{p}}(\lambda)\,\mathrm{d}\lambda} \tag{A7}$$

By applying the separability assumption for each wavelength, Eq. ($A$4), we get

$$\Psi(\theta) = \frac{\int V(\lambda)\,s(\lambda)\,E_{\mathrm{p}}(\lambda)\,\mathrm{d}\lambda}{\int V(\lambda)\,E_{\mathrm{p}}(\lambda)\,\mathrm{d}\lambda}\,\psi(\theta) \tag{A8}$$

so that the straylight parameter $s$ for light of arbitrary spectrum $E_{\mathrm{p}}(\lambda)$ is given by

$$s = \frac{\int V(\lambda)\,s(\lambda)\,E_{\mathrm{p}}(\lambda)\,\mathrm{d}\lambda}{\int V(\lambda)\,E_{\mathrm{p}}(\lambda)\,\mathrm{d}\lambda} \tag{A9}$$

We can adapt this formulation to our present problem by allowing the spectral irradiance on the pupil to change over time, $E_{\mathrm{p}}(\lambda, t)$, and defining the glare radiance indicator as $L_{\mathrm{GAN}}^{\mathrm{IOC}}(t) \equiv L_{\mathrm{eq}}(\theta_0, t)$ for an external glaring source located at an arbitrary angle $\theta_0$ within the visual field. Then, from Eqs. ($A$4) and ($A$5),

$$L_{\mathrm{GAN}}^{\mathrm{IOC}}(t) = K_{\mathrm{r}}\,\psi(\theta_0) \int V(\lambda)\,s(\lambda)\,E_{\mathrm{p}}(\lambda, t)\,\mathrm{d}\lambda \tag{A10}$$

Since the most intense glare effects are produced by the direct light from the lamps, the spectral glaring irradiance on the eye pupil $E_{\mathrm{p}}(\lambda, t)$ is proportional to the spectral radiant flux emitted per unit area by the artificial light sources surrounding the observer, $E(\lambda, t)$, appearing in Eq. (1). Then we can write $E_{\mathrm{p}}(\lambda, t) = C \cdot E(\lambda, t)$ in Eq. ($A$10), where $C$ is a constant, and, from Eq. (1) we get

$$F_\beta(\lambda, t) = F_\beta(\lambda) = C\,K_r\,\psi(\theta_0)\,V(\lambda)\,s(\lambda) \tag{A11}$$

which is independent from time. Hence, from Eq.(4),

$$\mathcal{H}_{\mathrm{GAN},k}^{\mathrm{IOC}} \equiv C\,K_r\,\psi(\theta_0) \int V(\lambda)\,s(\lambda)\,\widehat{\Phi}_k(\lambda)\,\mathrm{d}\lambda \tag{A12}$$

where $\widehat{\Phi}_k(\lambda)$ is the spectral radiant flux of a lamp of the $k$-th type, $k \in \{\mathrm{HPS}, \mathrm{LED}\}$, normalized to 1 lm emission. Finally, the ratio $\mathcal{H}_{\mathrm{GAN,LED}}^{\mathrm{IOC}}/\mathcal{H}_{\mathrm{GAN,HPS}}^{\mathrm{IOC}}$ is given by:

$$\frac{\mathcal{H}_{\mathrm{GAN,LED}}^{\mathrm{IOC}}}{\mathcal{H}_{\mathrm{GAN,HPS}}^{\mathrm{IOC}}} = \frac{\int V(\lambda)\,s(\lambda)\,\widehat{\Phi}_{\mathrm{LED}}(\lambda)\,\mathrm{d}\lambda}{\int V(\lambda)\,s(\lambda)\,\widehat{\Phi}_{\mathrm{HPS}}(\lambda)\,\mathrm{d}\lambda} \tag{A13}$$

Equations ($A$10)-($A$13), derived for photopic adaptation, can be rewritten for mesopic and scotopic observers by replacing the photopic efficacy $K_{\mathrm{r}}$ and the spectral sensitivity function $V(\lambda)$ by their mesopic and scotopic counterparts: The scotopic parameters are $K_{\mathrm{r}}' = 1700$ lm·W$^{-1}$ and $V'(\lambda)$, respectively [8], whereas the mesopic ones depend on the adaptation coefficient of the observer [31].